\documentclass[authoryear,review,11pt,a4paper]{elsarticle}

\usepackage{setspace}
\usepackage{fullpage}

\usepackage{color}

\usepackage{amsmath}

\DeclareMathOperator*{\argmin}{arg\,min}
\usepackage{amsfonts}
\usepackage{amssymb}
\usepackage{amsthm}

\usepackage{amscd}

\usepackage{bbm}

\usepackage{algorithm}
\usepackage{algpseudocode}

\usepackage{graphicx}

\usepackage{lscape}

\usepackage{booktabs}
\usepackage{multirow}
\usepackage{float}
\usepackage{caption}
\usepackage{subcaption}

\usepackage{lineno}

\usepackage{hyperref}

\usepackage{listings}
\lstnewenvironment{rc}[1][]{\lstset{language=R}}{}
\biboptions{authoryear}

\newdefinition{rmk}{Remark}
\newproof{pf}{Proof}

\journal{ }

\begin{document}

\begin{frontmatter}



\title{Optimal scheduling of interim analyses in group sequential trials}

\author[TGIUK,ICTU]{Zhangyi He\corref{cor1}\fnref{fn3}}
\ead{zhe1@georgeinstitute.org.uk}

\author[ICTU]{Suzie Cro}

\author[TGIAUS]{Laurent Billot}

\cortext[cor1]{Corresponding author.}
\fntext[fn3]{Affiliation at time of study}

\address[TGIUK]{The George Institute for Global Health, Imperial College London, London W12 7RZ, United Kingdom}
\address[ICTU]{Imperial Clinical Trials Unit, Imperial College London, London W12 7RH, United Kingdom}
\address[TGIAUS]{The George Institute for Global Health, University of New South Wales, Sydney 2042, Australia}

\begin{abstract}
Group sequential designs (GSDs) are well established and the most commonly used adaptive design in confirmatory clinical trials with interim analyses. However, they remain underutilised, and their implementation involves unique theoretical and practical decisions that demand careful consideration to optimise efficiency. A common practice is to schedule interim analyses at equal intervals based on calendar time or accumulated data. While straightforward, this approach does not completely exploit the potential sample size savings achievable with GSDs. To address this challenge, we develop \texttt{OptimInterim}, an \texttt{R}-based tool that can determine the optimal scheduling of interim analyses to minimise the expected sample size under the alternative hypothesis while controlling overall type I and type II errors. Our method accommodates trials with continuous or binary endpoints, allows multiple interim analyses and supports a range of stopping boundaries. Through extensive simulations, we demonstrate that optimally spaced interim analyses can yield substantial savings in expected sample size compared to equally spaced interim analyses, without compromising the maximum sample size, across various endpoint types, effect sizes, error rates and stopping rules. We illustrate its practical utility with two landmark trials evaluating steroid use in septic shock. Notably, for given type I and type II error rates, the optimal scheduling is independent of endpoint types and effect sizes, ensuring broad applicability across a wide range of trial contexts. To facilitate implementation, we offer a ready-to-use reference table of optimal schedules for up to eight interim analyses under commonly used error rates and stopping rules.
Access \texttt{OptimInterim} at \url{https://github.com/zhangyi-he/GSD_OptimInterim}.
\end{abstract}

\begin{keyword}
Confirmatory trial \sep
Group sequential design \sep
Interim analysis \sep
Optimal scheduling \sep
\texttt{R} implementation
\end{keyword}

\end{frontmatter}


\section{Introduction}
\label{sec:1}
Randomised controlled trials are regarded as the gold standard in clinical research, but these studies come with several shortcomings. For example, large sample sizes are usually required to be adequately powered, necessitating extended study durations. In response to these challenges, adaptive trial designs have emerged as a potential solution capable of improving the efficiency of clinical trials. Such designs introduce flexibility by allowing the trial's course to be modified during the trial based on accumulating results under prespecified rules. Compared to traditional fixed trial designs, adaptive trial designs are generally more efficient, informative and ethically sound, mainly because a smaller subject pool might be required \citep{pallmann2018}. Refer to \citet{chow2008}, \citet{kairalla2012} and \citet{bhatt2016} for excellent reviews of adaptive designs for clinical trials.

The group sequential design (GSD) is the most widely used adaptive design in clinical trials \citep{hatfield2016}, allowing a trial to stop early for efficacy or futility at preplanned interim analyses without inflating the overall type I error rate \citep{pocock1977,obrien1979,lan1983,kim1987,hwang1990}. Recently, \citet{stevely2015} conducted a systematic review of published group sequential trials and found that the majority (approximately 68\%) could be terminated early. Although the GSD has been well established in the literature \citep[see, \textit{e.g.},][for theoretical introduction]{jennison1999,wassmer2016}, it still involves a number of unique theoretical and practical choices that require careful consideration in its planning and optimisation (\textit{e.g.}, when to conduct interim analyses, what information to use and what statistical methods and stopping rules to apply). 
See \citet{meurer2021} for a detailed discussion.

Optimal GSDs have been developed in the literature to optimise the trial operating characteristics, mainly with respect to the stopping rule \citep[\textit{e.g.},][]{wang1987,eales1992,brittain1993,chang1996,barber2002,anderson2007,wason2012a,wason2012b,grayling2021}. Except for the stopping criteria, the timing of interim analyses is another important factor that can significantly influence the operating characteristics of the group sequential trial. 
\citet{brittain1993} introduced two- and three-stage GSDs that minimise the expected sample size (ESS) under a specified alternative hypothesis with respect to critical values and interim analysis timings for given type I and type II error rates. \citet{togo2013} studied how to schedule interim analyses to minimise the ESS under a specified alternative hypothesis in the Pocock design \citep{pocock1977} and the O’Brien-Fleming design \citep{obrien1979}, each implemented using the Lan-DeMets error spending approach \citep{lan1983}. They illustrated that, for a single interim analysis, the optimal timing is approximately 50\% of the planned subjects with available data when using the Pocock-type spending function and about 67\% when using the O’Brien-Fleming-type spending function. Their findings hold across different effect sizes for given type I and type II error rates but are limited to two-stage GSDs with normally distributed endpoints. For further discussion and guidance on optimal two-stage GSDs, see \citet{pilz2021} and \citet{lewis2023}.

A recent simulation study by \citet{li2023} illustrated that limiting group sequential trials to one or two interim analyses may be suboptimal, as it reduces the potential for early stopping, and increased efficiency can often be achieved with more interim analyses. 
Taking administrative feasibility into account, planning between four and eight interim analyses is generally considered practical \citep{todd2001}. \citet{xi2017} further investigated how to determine the optimal timing of interim analyses in multi-stage group sequential trials, but their focus was solely on futility analyses. 
The optimal scheduling of interim analyses in GSDs has not been thoroughly studied. In practice, interim analyses are typically planned at equally spaced intervals, whether defined by calendar time or accumulated information. However, this conventional strategy does not fully capitalise on the potential sample size savings offered by GSDs \citep{togo2013}.

To address this issue, we extend the work of \citet{togo2013} to determine the optimal timing of interim analyses in multi-stage GSDs, aiming to minimise the ESS under a specified alternative hypothesis while maintaining nominal type I and type II error rates. We introduce an \texttt{R}-based tool, \texttt{OptimInterim}, which can determine optimal interim analysis timings for group sequential trials with either continuous or binary endpoints. The tool allows for multiple interim analyses and various stopping boundaries. We assess the operating characteristics of GSDs with up to eight optimally spaced interim analyses across a range of endpoint types, effect sizes, error rates and stopping boundaries. To show the practical utility of \texttt{OptimInterim}, we redesign two landmark clinical studies on steroids use in septic shock: the HYdrocortisone for PREvention of Septic Shock (HYPRESS) trial \citep{keh2016} and the ADjunctive coRticosteroid trEatment iN criticAlly ilL patients with septic shock (ADRENAL) trial \citep{venkatesh2018}. We close by discussing possible further generalisations and extensions.
\section{Materials and Methods}
\label{sec:2}
In this section, we begin with a theoretical overview of GSDs and then formalise the problem of optimising the scheduling of interim analyses in terms of minimising the ESS under a specified alternative hypothesis, subject to predefined type I error rate ($\alpha$) and type II error rate ($\beta$). We present \texttt{OptimInterim}, an R-based tool designed to determine optimal interim analysis timings for a broad range of GSDs. We outline the trial scenarios considered for evaluating its operating characteristics and introduce two case studies that show its practical utility.

\subsection{Group sequential designs}
\label{sec:21}
We consider a group sequential trial with a total of $K>1$ analyses, comprising $K-1$ interim analyses and a final analysis, comparing a treatment to a control. Let $\delta$ represent the treatment effect, \textit{i.e.}, the difference between the treatment and the control, with $\delta>0$ indicating treatment superiority. Superiority is established by rejecting the null hypothesis $H_{0}:\delta=0$ in favour of the one-sided alternative hypothesis $H_{1}:\delta>0$. To present exact rather than asymptotic results, we assume that the primary outcome $Y$ is normally distributed with a known common variance $\sigma^{2}$ across both treatment groups. For simplicity, we further assume that recruitment is halted at each interim analysis and that all outcomes are immediately available. Extensions to scenarios with continuous enrolment and follow-up requirements are discussed in Section~\ref{sec:4}.


At the $k$-th analysis, we let $N_{0,k}$ and $N_{1,k}$ denote the cumulative sample sizes in the control and treatment groups. The natural estimator of the mean difference $\delta$ is the cumulative sample mean difference $\hat{\delta}_{k}=\bar{Y}_{1,k}-\bar{Y}_{0,k}$, where $\bar{Y}_{1,k}$ and $\bar{Y}_{0,k}$ are the cumulative sample means for the control and treatment arms at the $k$-th analysis. Given normal outcomes with common variance $\sigma^{2}$, the variance of $\hat{\delta}_{k}$ is
\begin{linenomath}
	\begin{equation*}
		\operatorname{Var}(\hat{\delta}_{k})
		=
		\sigma^{2} \biggl( \frac{1}{N_{1,k}}+\frac{1}{N_{0,k}} \biggr),
	\end{equation*}
\end{linenomath}
and the information at the $k$-th analysis is then defined as the reciprocal of this variance as
\begin{linenomath}
	\begin{equation}
		\label{eqn:2102}
		I_{k}
		=
		\frac{1}{\sigma^{2}} \frac{N_{1,k}N_{0,k}}{N_{1,k}+N_{0,k}}.
	\end{equation}
\end{linenomath}
The standardised test statistic $Z_{k}$ at the $k$-th analysis can then be expressed as 
\begin{linenomath}
	\begin{equation*}
		Z_{k}
		=
		\frac{\hat{\delta}_{k}}{\sqrt{\operatorname{Var}(\hat{\delta}_{k})}}
		=
		\hat{\delta}_{k}\sqrt{I_{k}}.
	\end{equation*}
\end{linenomath}
Under the independent increments property of the sufficient statistics, $\hat{\boldsymbol{\delta}}_{1:K}=(\hat{\delta}_{1},\hat{\delta}_{2},\ldots,\hat{\delta}_{K})^\intercal$, the standardised test statistics $\boldsymbol{Z}_{1:K}=(Z_{1},Z_{2},\ldots,Z_{K})^\intercal$ follows a multivariate normal distribution with mean $\mu_{k}=\delta\sqrt{I_{k}}$ for $k=1,2,\ldots,K$ and covariance $\Sigma_{kk'}=\sqrt{\min(I_{k},I_{k'})/\max(I_{k},I_{k'})}$ for $k,k'=1,2,\ldots,K$. This canonical joint distribution \citep{jennison1999} allows exact computation of joint stopping probabilities and precise calibration of stopping boundaries to maintain nominal type I error control.

At each interim analysis ($k < K$), the group sequential trial stops early for efficacy if $Z_{k} \geq u_{k}$ or for futility if $Z_{k} \leq l_{k}$ (binding or non-binding). If no boundary is crossed, the trial continues to the next analysis until the final analysis ($k=K$), where the null hypothesis $H_{0}$ is rejected if $Z_{K} \geq u_{K}$ and otherwise not rejected. The stopping boundary values are calibrated to maintain the control of the overall type I error rate at the desirable level \citep[see, \textit{e.g.},][]{pocock1977,obrien1979,lan1983,kim1987,hwang1990}.

For clarity, we denote by $\mathcal{C}_{\,\cdot\,,k}$ the continuation region and by $\mathcal{R}_{k}$ the rejection region at the $k$-th analysis, where $\mathcal{R}_{k}=\{z \mid z \geq u_{k}\}$. The overall type I error rate $\alpha$ can be expressed as 
\begin{linenomath}
	\begin{equation}
		\label{eqn:2104}
		\alpha 
		=
		\Pr(Z_{1} \in \mathcal{R}_{1} \mid H_{0})+\sum_{k=2}^{K} \Pr(\cap_{k'=1}^{k-1} \{Z_{k'} \in \mathcal{C}_{0,k'}\} \cap \{Z_{k} \in \mathcal{R}_{k}\} \mid H_{0}),
	\end{equation}
\end{linenomath}
where $\mathcal{C}_{0,k}=\{z \mid l_{k}<z<u_{k}\}$ for binding futility and $\mathcal{C}_{0,k}=\{z \mid z < u_{k}\}$ for non-binding or no futility. Similarly, the power $1 - \beta$ can be represented as 
\begin{linenomath}
	\begin{equation}
		\label{eqn:2105}
		1-\beta 
		=
		\Pr(Z_{1} \in \mathcal{R}_{1} \mid H_{1})+\sum_{k=2}^{K} \Pr(\cap_{k'=1}^{k-1} \{Z_{k'} \in \mathcal{C}_{1,k'}\} \cap \{Z_{k} \in \mathcal{R}_{k}\} \mid H_{1}),
	\end{equation}
\end{linenomath}
where $\mathcal{C}_{1,k}=\{z \mid l_{k}<z<u_{k}\}$ for binding or non-binding futility and $\mathcal{C}_{1,k}=\{z \mid z<u_{k}\}$ for no futility. For consistency, we let $l_{K}=u_{K}$. The probabilities in Eqs.~(\ref{eqn:2104}) and (\ref{eqn:2105}) are evaluated through the multivariate normal distribution defined above. For a two-sided alternative hypothesis $H_{1}:\delta \neq 0$, the continuation and rejection regions are adjusted accordingly to accommodate both tails of the distribution.

\subsection{Optimal scheduling of interim analyses}
\label{sec:22}
Let $N_{k}=N_{0,k}+N_{1,k}$ denote the cumulative total sample size at the $k$-th analysis. Minimising the ESS under a given alternative hypothesis, while controlling overall type I and type II errors, can then be expressed as a constrained optimisation problem over both the stopping boundaries and the scheduling of interim analyses, formulated as
\begin{linenomath}
	\begin{equation}
		\label{eqn:2201}
		\begin{aligned}
			& \underset{\boldsymbol{N}_{1:K},\boldsymbol{u}_{1:K},\boldsymbol{l}_{1:K}}{\text{minimise}}
			& & \operatorname{E}(N \mid H_{1}) \\
			& \text{subject to}
			& & \Pr(Z_{1} \in \mathcal{R}_{1} \mid H_{0})+\sum_{k=2}^{K} \Pr(\cap_{k'=1}^{k-1} \{Z_{k'} \in \mathcal{C}_{0,k'}\} \cap \{Z_{k} \in \mathcal{R}_{k}\} \mid H_{0}) \leq \alpha \\
			&
			& & \Pr(Z_{1} \in \mathcal{R}_{1} \mid H_{1})+\sum_{k=2}^{K} \Pr(\cap_{k'=1}^{k-1} \{Z_{k'} \in \mathcal{C}_{1,k'}\} \cap \{Z_{k} \in \mathcal{R}_{k}\} \mid H_{1}) \geq 1-\beta \\
			&
			& & N_{1}<N_{2}<\ldots<N_{K} \text{ with } N_{k} \in \mathbb{N}^{+} \text{ for } k=1,2,\ldots,K,
		\end{aligned}
	\end{equation}
\end{linenomath}
where 
\begin{linenomath}
	\begin{equation}
		\label{eqn:2202}
		\operatorname{E}(N \mid H_{1})
		=
		N_{1}\Pr(Z_{1} \in \mathcal{C}_{1,1}^\complement \mid H_{1}) + \sum_{k=2}^{K}N_{k}\Pr(\cap_{k'=1}^{k-1} \{Z_{k'} \in \mathcal{C}_{1,k'}\} \cap \{Z_{k} \in \mathcal{C}_{1,k}^\complement\} \mid H_{1})	
	\end{equation}
\end{linenomath}
with $\mathcal{C}_{1,k}^\complement=\mathbb{R} \setminus \mathcal{C}_{1,k}$ for $k=1,2\ldots,K$ and $\mathcal{C}_{1,K}=\emptyset$.

As stopping rules in group sequential trials are typically constructed by established methods \citep[\textit{e.g.},][]{pocock1977,obrien1979,lan1983,kim1987,hwang1990}, the optimisation problem in Eq.~(\ref{eqn:2201}) reduces to one that focuses solely on the scheduling of interim analyses. More specifically, we introduce the information rate $t_{k}=I_{k}/I_{K}$ for $k=1,2,\ldots,K$, which quantifies trial progress as a proportion of total information, reaching $t_{K}=1$ at maximum information. For an allocation ratio $r$, the information $I_{k}$ defined in Eq.~(\ref{eqn:2102}) simplifies to
\begin{linenomath}
	\begin{equation*}
		I_{k}
		=
		\frac{1}{\sigma^{2}} \frac{r}{(1+r)^{2}}N_{k}
	\end{equation*}
\end{linenomath}
for $k=1,2,\ldots,K$, therefore reducing the information rate to $t_{k}=N_{k}/N_{K}$. Under a specified stopping rule, the information rates $\boldsymbol{t}_{1:K}=(t_{1},t_{2},\ldots,t_{K})^\intercal$ fully determine the allocation of type I and type II error across analyses, denoted $\alpha_{k}$ and $\beta_{k}$ at the $k$-th analysis, with $\sum_{k=1}^{K}\alpha_{k}=\alpha$ and $\sum_{k=1}^{K}\beta_{k}=\beta$. Given $\boldsymbol{\alpha}_{1:K}$ and $\boldsymbol{\beta}_{1:K}$, the stopping boundaries $\boldsymbol{u}_{1:K}$ and $\boldsymbol{l}_{1:K}$, together with the maximum sample size $N_{K}$, can then be uniquely determined through
\begin{linenomath}
	\begin{align*}
		\alpha_{1} 
		&= 
		\Pr(Z_{1} \in \mathcal{R}_{1} \mid H_{0}) \notag \\
		&=
		F(\mathcal{R}_{1} \mid 0,\Sigma_{11}) \\
		\alpha_{k} 
		&= 
		\Pr(\cap_{k'=1}^{k-1} \{Z_{k'} \in \mathcal{C}_{0,k'}\} \cap \{Z_{k} \in \mathcal{R}_{k}\} \mid H_{0}) \notag \\
		&=
		F(\mathcal{C}_{0,1} \times \mathcal{C}_{0,2} \times \cdots \times \mathcal{C}_{0,k-1} \times \mathcal{R}_{k} \mid \boldsymbol{0}_{1:k},\boldsymbol{\Sigma}_{k \times k}) \\
		\beta_{1} 
		&= 
		\Pr(Z_{1} \in (\mathcal{R}_{1} \cup \mathcal{C}_{1,1})^\complement \mid H_{1}) \notag \\
		&= 
		F((\mathcal{R}_{1} \cup \mathcal{C}_{1,1})^\complement \mid \mu_{1},\Sigma_{11}) \\
		\beta_{k} 
		&= 
		\Pr(\cap_{k'=1}^{k-1} \{Z_{k'} \in \mathcal{C}_{1,k'}\} \cap \{Z_{k} \in (\mathcal{R}_{k} \cup \mathcal{C}_{1,k})^\complement \} \mid H_{1}) \notag \\
		&=
		F(\mathcal{C}_{1,1} \times \mathcal{C}_{1,2} \times \cdots \times \mathcal{C}_{1,k-1} \times (\mathcal{R}_{k} \cup \mathcal{C}_{1,k})^\complement \mid \boldsymbol{\mu}_{1:k},\boldsymbol{\Sigma}_{k \times k})
	\end{align*}
\end{linenomath}
for $k=2,\ldots,K$, where $F(\,\cdot\, \mid \boldsymbol{\mu}_{1:k},\boldsymbol{\Sigma}_{k \times k})$ represents a $k$-dimensional normal distribution with mean $\mu_{k'}=\theta\sqrt{t_{k'}/t_{1}}$ for $k'=1,2,\ldots,k$ and covariance $\Sigma_{k'k''}=\sqrt{\min(t_{k'},t_{k''})/\max(t_{k'},t_{k''})}$ for $k',k''=1,2,\ldots,k$. Here, $\theta$ denotes the canonical drift parameter, derived via the recursive formulation of \citet{armitage1969} and determined with $\boldsymbol{t}_{1:K}$, $\boldsymbol{\alpha}_{1:K}$ and $\boldsymbol{\beta}_{1:K}$. This parameter captures the noncentrality under the alternative hypothesis and scales the expected trajectory of the test statistics across analyses according to the information rate.

Given $\mu_{k}=\theta\sqrt{t_{k}/t_{1}}=\delta^{*}\sqrt{I_{k}}$, the maximum sample size $N_{K}$ can be written down as
\begin{linenomath}
	\begin{equation}
		\label{eqn:2208}
		N_{K}
		=
		\biggl( \frac{\sigma}{\delta^{*}} \biggr)^{2} \frac{(1+r)^{2}}{r} \frac{\theta^{2}}{t_{1}},
	\end{equation}
\end{linenomath}
where $\delta^{*}$ represents the target effect size under $H_{1}$. With Eq.~(\ref{eqn:2208}), we can reformulate $\mathbb{E}(N \mid H_{1})$ in Eq.~(\ref{eqn:2202}) as
\begin{linenomath}
	\begin{equation}
		\label{eqn:2209}
		\mathbb{E}(N \mid H_{1})
		=
		\frac{N_{0}\theta^{2}}{(z_{1-\alpha}+z_{1-\beta})^{2}} \biggl(1 + \sum_{k=2}^{K}\frac{t_{k}-t_{k-1}}{t_{1}}F(\mathcal{C}_{1,1} \times \mathcal{C}_{1,2} \times \cdots \times \mathcal{C}_{1,k-1} \mid \boldsymbol{\mu}_{1:k},\boldsymbol{\Sigma}_{k \times k}) \biggr),
	\end{equation}
\end{linenomath}
where 
\begin{linenomath}
	\begin{equation}
		\label{eqn:2210}
		N_{0}
		=
		\biggl( \frac{\sigma}{\delta^{*}} \biggr)^{2} \frac{(1+r)^{2}}{r} (z_{1-\alpha}+z_{1-\beta})^{2}
	\end{equation}
\end{linenomath}
is the sample size required for the corresponding fixed design with $z_{p}$ denoting the $p$-th quantile of the standard normal distribution. Hence, for a prespecified stopping rule that ensures control of type I and type II error rates, the optimisation problem in Eq.~(\ref{eqn:2201}) reduces to one solely over the scheduling of interim analyses, formulated as
\begin{linenomath}
	\begin{equation}
		\label{eqn:2211}
		\begin{aligned}
			& \underset{\boldsymbol{t}_{1:K-1}}{\text{minimise}}
			& & \frac{N_{0}\theta^{2}}{(z_{1-\alpha}+z_{1-\beta})^{2}} \biggl(1 + \sum_{k=2}^{K}\frac{t_{k}-t_{k-1}}{t_{1}}F(\mathcal{C}_{1,1} \times \mathcal{C}_{1,2} \times \cdots \times \mathcal{C}_{1,k-1} \mid \boldsymbol{\mu}_{1:k},\boldsymbol{\Sigma}_{k \times k}) \biggr) \\
			& \text{subject to}
			& & 0<t_{1}<t_{2}<\cdots<t_{K-1}<1.
		\end{aligned}
	\end{equation}
\end{linenomath}
Notably, from Eq.~(\ref{eqn:2211}), the optimal scheduling of interim analyses 
\begin{linenomath}
	\begin{equation*}
		\hat{\boldsymbol{t}}_{1:K-1}
		=
		\argmin_{0<t_{1}<t_{2}<\cdots<t_{K-1}<1} \theta^{2} \biggl(1 + \sum_{k=2}^{K}\frac{t_{k}-t_{k-1}}{t_{1}}F(\mathcal{C}_{1,1} \times \mathcal{C}_{1,2} \times \cdots \times \mathcal{C}_{1,k-1} \mid \boldsymbol{\mu}_{1:k},\boldsymbol{\Sigma}_{k \times k}) \biggr)
	\end{equation*}
\end{linenomath}
is independent of the standardised effect size $\delta^{*}/\sigma$. These results are equally applicable to test statistics that are asymptotically normally distributed, \textit{e.g.}, those arising from binary endpoints.

\subsection{\texttt{OptimInterim}}
\label{sec:23}
We introduce \texttt{OptimInterim}, an \texttt{R}-based tool for determining the optimal timing of interim analyses in GSDs, aimed at minimising the ESS under $H_{1}$ while controlling type I and type II error rates. \texttt{OptimInterim} offers substantial flexibility, accommodating a wide range of stopping rules for efficacy and/or futility, including classical stopping boundaries \citep[\textit{e.g.},][]{haybittle1971,peto1976,pocock1977,obrien1979,wang1987,pampallona1994} and commonly used alpha- and beta-spending functions \citep[\textit{e.g.},][]{lan1983,kim1987,hwang1990}, as well as fully customisable spending functions. This tool supports both one- and two-sided tests and allows for either binding or non-binding futility stopping rules. 

Our implementation of \texttt{OptimInterim} is based on the \texttt{rpact} package \citep{wassmer2024}, a comprehensive and validated \texttt{R} package for the design, simulation and analysis of confirmatory adaptive clinical trials. More specifically, the calculation of the ESS under $H_{1}$ in \texttt{OptimInterim} directly utilises the core functionality of \texttt{rpact}, ensuring numerical accuracy and full consistency with methodologies implemented in established GSD software tools. To identify optimal interim analysis timings, the minimisation of the ESS under $H_{1}$ is carried out through the Nelder-Mead algorithm \citep{nelder1965}, which is executed via the \texttt{optimx} function from the \texttt{optimx} package \citep{nash2023}. Given that the Nelder-Mead algorithm conducts a restricted global search, which may result in convergence to a local minimum rather than the global optimum, we mitigate this risk by repeating the ESS minimisation under $H_{1}$ from a range of systematically varied starting values for the information rates, therefore enabling broader coverage of the search space. After each run, the ESS values are compared, and the process is iterated until no further improvement in the ESS is observed, reducing the risk of convergence to suboptimal solutions and increasing confidence that the resulting interim analysis schedule is close to the true global optimum.

\texttt{OptimInterim} currently provides a single function, \texttt{getOptimalInformationRates}, which produces the optimal information rates for interim analyses in two-arm group sequential trials with continuous or binary endpoints in terms of minimising the ESS under $H_{1}$ while maintaining the control of type I and type II error rates. The package is available at \url{https://github.com/zhangyi-he/GSD_OptimInterim}, with complete documentation for all arguments and multiple illustrative examples. The full design, simulation and evaluation of GSDs are carried out through \texttt{rpact}, with \texttt{OptimInterim} providing the optimised interim analysis schedule as input.

\subsection{Performance evaluation}
\label{sec:25}
To assess the performance of \texttt{OptimInterim} in group sequential trials, we consider two-arm superiority trials under balanced randomisation, comparing a treatment to a control. Suppose that the response of each participant is independent and identically normally distributed with a known common standard deviation $\sigma=1$. We test the null hypothesis $H_{0}:\delta \leq 0$ against the alternative hypothesis $H_{1}:\delta>0$ with a design effect size of $\delta^{*}=0.5$. We fix the type I error rate at $\alpha=0.025$ for one-sided tests and at $\alpha=0.05$ for two-sided tests while varying the type II error rate over $\beta \in \{0.1,0.2\}$, corresponding to 90\% and 80\% power, respectively, which are widely adopted in practice.

For each combination of type I and type II error rates, we evaluate the operating characteristics of three widely used GSDs: the Haybittle-Peto design \citep{haybittle1971,peto1976}, the Pocock design \citep{pocock1977} and the O'Brien-Fleming \citep{obrien1979} design, all of which allow early stopping for efficacy. 
Unless otherwise specified, stopping boundaries for the Pocock and O'Brien-Fleming designs are approximated through the alpha spending function \citep{lan1983} to mimic their original stopping rules.
For each design, we consider scenarios with up to eight optimally spaced interim analyses (corresponding to a total of nine trial stages, including the final analysis) generated by \texttt{OptimInterim}. For comparison, we also report the operating characteristics of the same designs when interim analyses are equally scheduled, a common practice in trial planning.

The operating characteristics of interests are the MSS and ESSs under $H_{0}$, $H_{0}/H_{1}$ and $H_{1}$, where $H_{0}/H_{1}$ denotes a target effect size of $\delta=\delta^{*}/2$. As the inflation factor is independent of the standardised effect size (see Eqs.~(\ref{eqn:2208})--(\ref{eqn:2210})), we instead present the maximum inflation factors (MIFs) under $H_{0}$, $H_{0}/H_{1}$ and $H_{1}$, as well as the expected inflation factors (EIFs) under each of these hypotheses. Here the EIF is defined as the ratio of the ESS under a given hypothesis to the sample size required in the corresponding fixed design, and the MIF is defined analogously using the MSS. As shown in the previous section, the optimal timing of interim analyses is also independent of the standardised effect size for a given stopping rule with predefined type I and type II error rates, ensuring that our conclusions remain valid across different effect sizes.

Our performance evaluation focuses on a continuous primary endpoint; however, the findings and conclusions are equally applicable to binary endpoints, as the corresponding test statistics are asymptotically normally distributed.

\subsection{Case studies}
\label{sec:24}
We show the practical use of \texttt{OptimInterim} through the HYPRESS trial \citep{keh2016}, which compared hydrocortisone with placebo in patients with severe sepsis, and the ADRENAL trial \citep{venkatesh2018}, which evaluated the same comparison in patients with septic shock. Both trials were originally conducted through GSDs but did not stop early in practice. Using the \texttt{rpact} package \citep{wassmer2024} with \texttt{OptimInterim}, we redesign each trial to incorporate optimally spaced interim analyses and compare the resulting operating characteristics with those of the original designs. All other trial settings are kept identical to their originals, with the sole modification being the scheduling of interim analyses.

\subsubsection{The HYPRESS trial}
\label{sec:241}
The HYPRESS trial was a multicentre, placebo-controlled, double-blind, randomised clinical trial designed to test the primary hypothesis that hydrocortisone, compared to placebo, reduces progression to septic shock in patients with sepsis. The primary endpoint was the occurrence of septic shock within 14 days. The design assumed a 40\% 14-day occurrence of septic shock in the placebo group and aimed to detect a clinically meaningful reduction of 15\% in the hydrocortisone group, with a two-sided type I error rate of 0.05 and a type II error rate of 0.2. HYPRESS used a three-stage O'Brien-Fleming GSD \citep{obrien1979} with balanced randomisation, incorporating two interim analyses for efficacy conducted at 33.3\% and 66.7\% of the planned total enrolment with available primary outcome data. The stopping boundary was determined by the alpha-spending function of \citet{lan1983}. Further details of the HYPRESS trial, including its study protocol, are provided in \citet{keh2016}.

\subsubsection{The ADRENAL trial}
\label{sec:242}
The ADRENAL trial was a multicentre, placebo-controlled, double-blind, randomised clinical trial conducted to examine the primary hypothesis that hydrocortisone, compared to placebo, decreases 90-day all-cause mortality in patients admitted to an intensive care unit with septic shock. The primary endpoint was all-cause mortality 90 days after randomisation. Given a 33\% 90-day all-cause mortality in patients with septic shock receiving placebo, ADRENAL aimed to establish a clinically meaningful 5\% reduction in those receiving hydrocortisone, with a two-sided type I error rate of 0.05 and a type II error rate of 0.1. ADRENAL employed a three-stage Haybittle-Peto GSD \citep{haybittle1971,peto1976} with equal allocation and two interim analyses conducted to evaluate efficacy when primary outcome data were available for 950 and 2500 patients (\textit{i.e.}, at the information rate of 25.0\% and 65.8\%, respectively). Further details of the ADRENAL trial, including the trial protocol and statistical analysis plan, are provided in \citet{venkatesh2018}.

\section{Results}
\label{sec:3}
This section evaluates the operating characteristics of GSDs with optimally spaced interim analyses across a range of scenarios, including the HYPRESS and ADRENAL trials. For clarity, unless otherwise noted, optimal design refers to a GSD with optimally spaced interim analyses, while conventional design refers to one with equally spaced interim analyses.

\subsection{Operating characteristics}
\label{sec:31}
In the following, we present the operating characteristics of the Haybittle-Peto, Pocock and O'Brien-Fleming designs, each incorporating up to eight optimally spaced interim analyses, for group sequential trials with a one-sided type I error rate of 0.025. For trials with a two-sided type I error rate of 0.05, the optimal interim analysis timings are identical, and the operating characteristics are similarly comparable.

\subsubsection{Haybittle-Peto group sequential design}
\label{sec:311}
We show the optimal information rates for the Haybittle-Peto design with up to nine stages (eight interim analyses plus a final analysis) for one-sided type I error of 0.025 and type II error of 0.1 and 0.2 in Table~\ref{tab:311} and illustrate their MIFs and EIFs under $H_{0}$, $H_{0}/H_{1}$ and $H_{1}$ in Figure~\ref{fig:311}, respectively (also summarised in Supplemental Material, Table~\href{run:./ZH2023_Supplemental_Material.pdf}{S1}). The stopping boundaries for the optimal Haybittle-Peto design can be found in Supplemental Material, Figure~\href{run:./ZH2023_Supplemental_Material.pdf}{S1}.

\begin{table}[!ht]
	\centering
	\begin{tabular}{cccccccccccc}
		\toprule
		                             & Stage & 1st             & 2nd             & 3rd             & 4th             & 5th             & 6th             & 7th             & 8th             & 9th   \\
		\hline
		\multirow{9}{*}{$\beta=0.1$} & 1     & 100.0           &                 &                 &                 &                 &                 &                 &                 &       \\
		                             & 2     & \phantom{0}59.4 & 100.0           &                 &                 &                 &                 &                 &                 &       \\
		                             & 3     & \phantom{0}44.4 & \phantom{0}70.4 & 100.0           &                 &                 &                 &                 &                 &       \\
		                             & 4     & \phantom{0}36.2 & \phantom{0}56.1 & \phantom{0}76.2 & 100.0           &                 &                 &                 &                 &       \\
		                             & 5     & \phantom{0}30.8 & \phantom{0}47.2 & \phantom{0}63.0 & \phantom{0}79.8 & 100.0           &                 &                 &                 &       \\
		                             & 6     & \phantom{0}27.1 & \phantom{0}41.1 & \phantom{0}54.3 & \phantom{0}67.8 & \phantom{0}82.3 & 100.0           &                 &                 &       \\
		                             & 7     & \phantom{0}24.3 & \phantom{0}36.6 & \phantom{0}48.0 & \phantom{0}59.4 & \phantom{0}71.3 & \phantom{0}84.2 & 100.0           &                 &       \\
		                             & 8     & \phantom{0}22.2 & \phantom{0}33.1 & \phantom{0}43.2 & \phantom{0}53.2 & \phantom{0}63.3 & \phantom{0}74.0 & \phantom{0}85.7 & 100.0           &       \\
		                             & 9     & \phantom{0}20.7 & \phantom{0}30.7 & \phantom{0}39.7 & \phantom{0}48.6 & \phantom{0}57.7 & \phantom{0}66.6 & \phantom{0}76.4 & \phantom{0}86.9 & 100.0 \\
		\hline
		\multirow{9}{*}{$\beta=0.2$} & 1     & 100.0           &                 &                 &                 &                 &                 &                 &                 &       \\
		                             & 2     & \phantom{0}61.2 & 100.0           &                 &                 &                 &                 &                 &                 &       \\
		                             & 3     & \phantom{0}46.6 & \phantom{0}72.1 & 100.0           &                 &                 &                 &                 &                 &       \\
		                             & 4     & \phantom{0}38.6 & \phantom{0}58.4 & \phantom{0}77.7 & 100.0           &                 &                 &                 &                 &       \\
		                             & 5     & \phantom{0}33.5 & \phantom{0}50.0 & \phantom{0}65.4 & \phantom{0}81.2 & 100.0           &                 &                 &                 &       \\
		                             & 6     & \phantom{0}30.1 & \phantom{0}44.2 & \phantom{0}57.2 & \phantom{0}70.1 & \phantom{0}83.7 & 100.0           &                 &                 &       \\
		                             & 7     & \phantom{0}27.6 & \phantom{0}40.1 & \phantom{0}51.4 & \phantom{0}62.4 & \phantom{0}73.6 & \phantom{0}85.5 & 100.0           &                 &       \\
		                             & 8     & \phantom{0}25.7 & \phantom{0}36.9 & \phantom{0}46.9 & \phantom{0}56.6 & \phantom{0}66.3 & \phantom{0}76.3 & \phantom{0}87.0 & 100.0           &       \\
		                             & 9     & \phantom{0}24.4 & \phantom{0}34.6 & \phantom{0}43.7 & \phantom{0}52.3 & \phantom{0}60.9 & \phantom{0}69.5 & \phantom{0}78.5 & \phantom{0}88.1 & 100.0 \\
		\bottomrule                                   
	\end{tabular}%
	\caption{Information rates (\%) for the optimal Haybittle-Peto design incorporating up to eight interim analyses for efficacy with a one-sided type I error rate of 0.025.}
	\label{tab:311}
\end{table}

\begin{figure}[!ht]
	\centering
	\includegraphics[width=\linewidth]{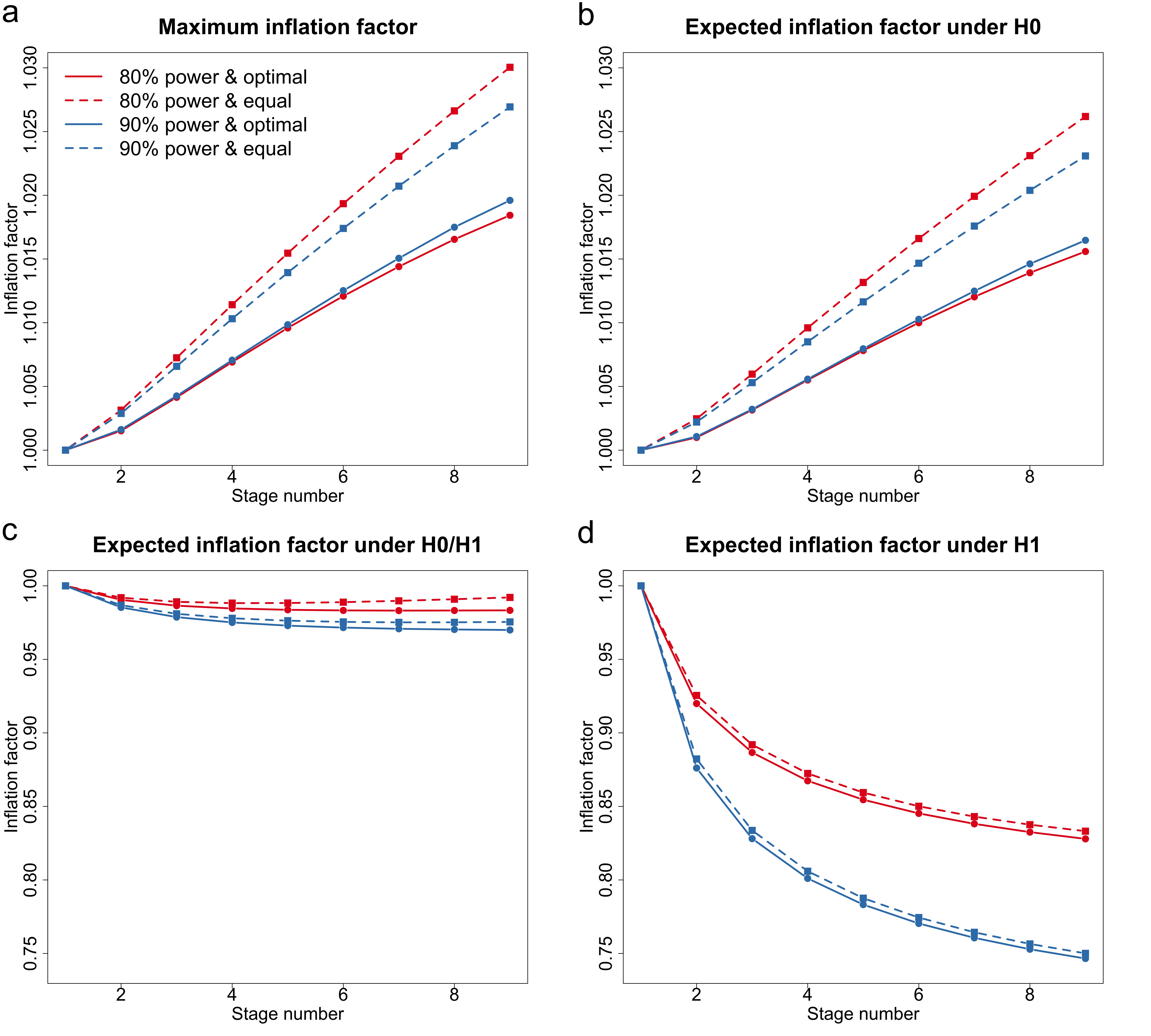}
	\caption{Operating characteristics for the optimal Haybittle-Peto design incorporating up to eight interim analyses for efficacy with a one-sided type I error rate of 0.025.}
	\label{fig:311}
\end{figure}

As shown in Figure~\ref{fig:311}, the inflation factors exhibit distinct trends with the number of stages. Both the MIF and the EIF under $H_{0}$ increase almost linearly and remain above 1 across stage numbers, although the increase is modest in magnitude. 
The EIF under $H_{0}/H_{1}$ remains close to but slightly below 1 with only a mild downward trend across stage numbers. The difference between optimal and conventional designs is generally minor, but becomes more apparent with more stages, where optimal spacing consistently achieves marginally lower inflation. By contrast, the EIF under $H_{1}$ falls steeply as the number of stages rises, particularly from one to four stages. 

The Haybittle-Peto design achieves sample size savings when a true treatment effect exists, with the magnitude of the benefit increasing as the effect size increases, although at the expense of requiring more patients when no effect is present. Across all stage numbers and power levels, optimally scheduled interim analyses consistently outperform equal spacing, though the margin is modest. This advantage is slightly greater at lower power, where reductions in inflation costs and gains in trial efficiency are more evident when moving from equal to optimal spacing.
\subsubsection{Pocock group sequential design}
\label{sec:312}
For the Pocock design, we list the optimal information rates for trials with up to nine stages in Table~\ref{tab:312}, along with the plots in Figure~\ref{fig:312} separately illustrating their MIFs and EIFs under $H_{0}$, $H_{0}/H_{1}$ and $H_{1}$. See Supplemental Material, Table~\href{run:./ZH2023_Supplemental_Material.pdf}{S2} and Figure~\href{run:./ZH2023_Supplemental_Material.pdf}{S2} for their corresponding operating characteristics and stopping boundaries.

\begin{table}[!ht]
	\centering
	\begin{tabular}{cccccccccccc}
		\toprule
		                             & Stage & 1st             & 2nd             & 3rd             & 4th             & 5th             & 6th             & 7th             & 8th             & 9th   \\
		\hline
		\multirow{9}{*}{$\beta=0.1$} & 1     & 100.0           &                 &                 &                 &                 &                 &                 &                 &       \\
		                             & 2     & \phantom{0}48.4 & 100.0           &                 &                 &                 &                 &                 &                 &       \\
		                             & 3     & \phantom{0}35.3 & \phantom{0}64.1 & 100.0           &                 &                 &                 &                 &                 &       \\
		                             & 4     & \phantom{0}29.3 & \phantom{0}50.2 & \phantom{0}71.9 & 100.0           &                 &                 &                 &                 &       \\
		                             & 5     & \phantom{0}25.9 & \phantom{0}42.6 & \phantom{0}58.7 & \phantom{0}76.5 & 100.0           &                 &                 &                 &       \\
		                             & 6     & \phantom{0}23.6 & \phantom{0}37.8 & \phantom{0}50.8 & \phantom{0}64.3 & \phantom{0}79.7 & 100.0           &                 &                 &       \\
		                             & 7     & \phantom{0}22.1 & \phantom{0}34.5 & \phantom{0}45.6 & \phantom{0}56.6 & \phantom{0}68.4 & \phantom{0}81.9 & 100.0           &                 &       \\
		                             & 8     & \phantom{0}20.9 & \phantom{0}32.1 & \phantom{0}41.7 & \phantom{0}51.1 & \phantom{0}60.9 & \phantom{0}71.4 & \phantom{0}83.6 & 100.0           &       \\
		                             & 9     & \phantom{0}19.9 & \phantom{0}29.9 & \phantom{0}38.2 & \phantom{0}46.2 & \phantom{0}54.6 & \phantom{0}63.5 & \phantom{0}73.2 & \phantom{0}84.6 & 100.0 \\
		\hline
		\multirow{9}{*}{$\beta=0.2$} & 1     & 100.0           &                 &                 &                 &                 &                 &                 &                 &       \\
		                             & 2     & \phantom{0}51.0 & 100.0           &                 &                 &                 &                 &                 &                 &       \\
		                             & 3     & \phantom{0}38.2 & \phantom{0}66.8 & 100.0           &                 &                 &                 &                 &                 &       \\
		                             & 4     & \phantom{0}32.3 & \phantom{0}53.4 & \phantom{0}74.2 & 100.0           &                 &                 &                 &                 &       \\
		                             & 5     & \phantom{0}28.9 & \phantom{0}46.1 & \phantom{0}61.8 & \phantom{0}78.5 & 100.0           &                 &                 &                 &       \\
		                             & 6     & \phantom{0}26.8 & \phantom{0}41.5 & \phantom{0}54.3 & \phantom{0}67.2 & \phantom{0}81.3 & 100.0           &                 &                 &       \\
		                             & 7     & \phantom{0}25.3 & \phantom{0}38.3 & \phantom{0}49.3 & \phantom{0}59.9 & \phantom{0}71.0 & \phantom{0}83.3 & 100.0           &                 &       \\
		                             & 8     & \phantom{0}24.2 & \phantom{0}35.9 & \phantom{0}45.6 & \phantom{0}54.8 & \phantom{0}64.0 & \phantom{0}73.7 & \phantom{0}84.7 & 100.0           &       \\
		                             & 9     & \phantom{0}23.4 & \phantom{0}34.1 & \phantom{0}42.8 & \phantom{0}50.9 & \phantom{0}58.9 & \phantom{0}67.1 & \phantom{0}75.9 & \phantom{0}85.8 & 100.0 \\
		\bottomrule                                   
	\end{tabular}%
	\caption{Information rates (\%) for the optimal Pocock design incorporating up to eight interim analyses for efficacy with a one-sided type I error rate of 0.025.}
	\label{tab:312}
\end{table}

\begin{figure}[!ht]
	\centering
	\includegraphics[width=\linewidth]{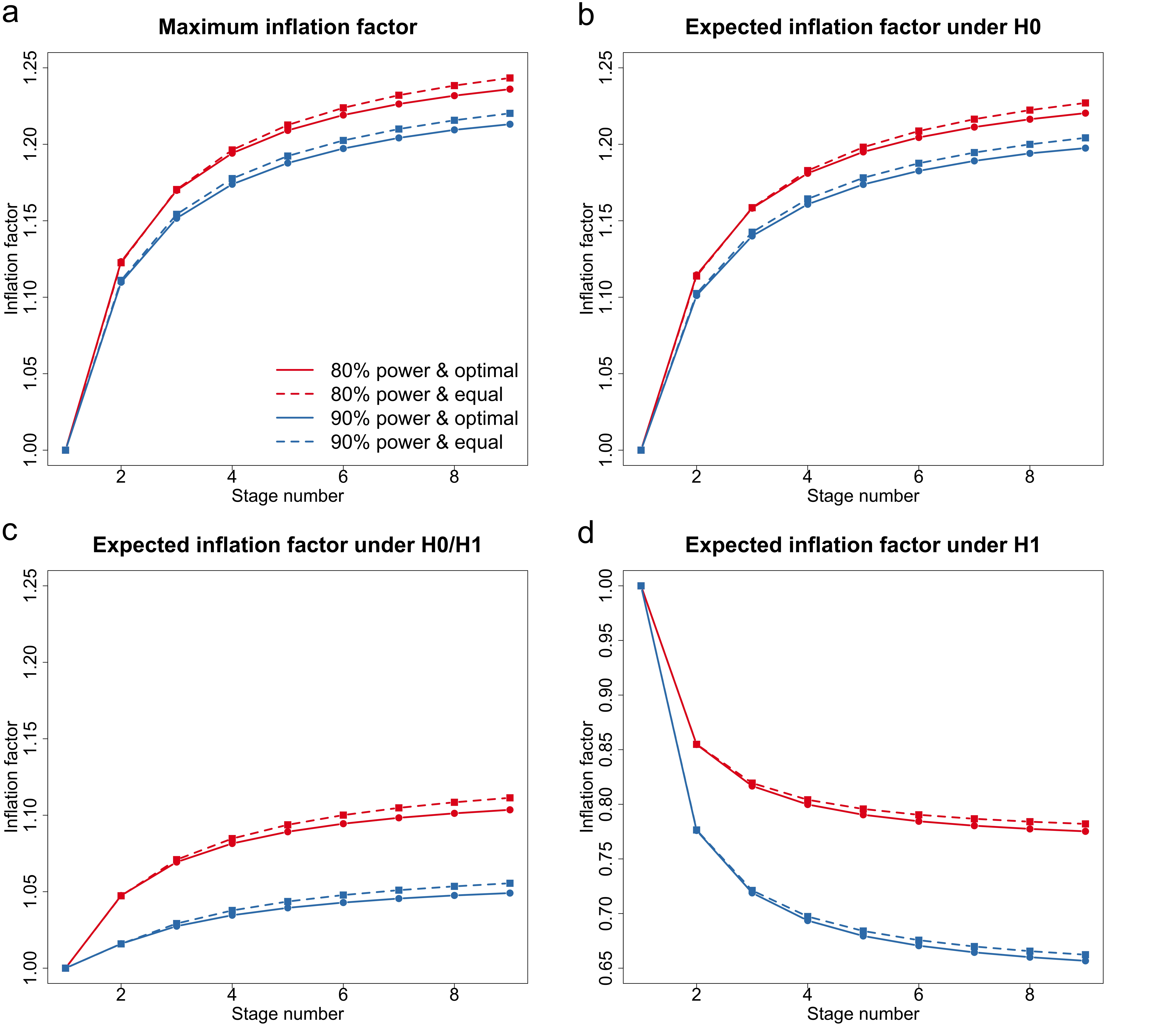}
	\caption{Operating characteristics for the optimal Pocock design incorporating up to eight interim analyses for efficacy with a one-sided type I error rate of 0.025.}
	\label{fig:312}
\end{figure}

We observe from Figure~\ref{fig:312} that for the MIF and the EIF under $H_{0}$, inflation increases sharply when moving from two to four total stages, after which the rate of increase slows, highlighting diminishing inflation cost for additional stages beyond four. Under $H_{0}/H_{1}$, inflation also grows with additional stages but at a smaller magnitude than under $H_{0}$. Across these settings, optimal designs consistently achieve slightly lower inflation factors than conventional designs, with the gap becoming more noticeable as the number of stages increases. In contrast, under $H_{1}$, inflation decreases sharply as the number of stages increases, particularly between two and four stages, after which the rate of reduction slows, reflecting diminishing efficiency gains from adding further interim analyses beyond four.


The Pocock design generally requires a larger sample size when the treatment effect is absent or smaller than the target, but achieves substantial savings when the assumed target effect is realised. Within this design, optimal scheduling provides a small yet consistent advantage over equal spacing across all stage numbers and power levels. The benefit becomes more evident as the number of stages or the required power increases, although it remains negligible overall.



\subsubsection{O'Brien-Fleming group sequential design}
\label{sec:313}
For the O'Brien-Fleming design, we report the optimal information rates for trials with up to nine stages in Table~\ref{tab:313}. The corresponding stopping boundaries can be found in Supplemental Material, Figure~\href{run:./ZH2023_Supplemental_Material.pdf}{S3}. We illustrate their MIFs and EIFs under $H_{0}$, $H_{0}/H_{1}$ and $H_{1}$ in Figure~\ref{fig:313}, summarised in Supplemental Material, Table~\href{run:./ZH2023_Supplemental_Material.pdf}{S3}.

\begin{table}[!ht]
	\centering
	\begin{tabular}{cccccccccccc}
		\toprule
		                             & Stage & 1st             & 2nd             & 3rd             & 4th             & 5th             & 6th             & 7th             & 8th             & 9th   \\
		\hline
		\multirow{9}{*}{$\beta=0.1$} & 1     & 100.0           &                 &                 &                 &                 &                 &                 &                 &       \\
		                             & 2     & \phantom{0}65.7 & 100.0           &                 &                 &                 &                 &                 &                 &       \\
		                             & 3     & \phantom{0}54.9 & \phantom{0}74.2 & 100.0           &                 &                 &                 &                 &                 &       \\
		                             & 4     & \phantom{0}49.3 & \phantom{0}63.4 & \phantom{0}78.6 & 100.0           &                 &                 &                 &                 &       \\
		                             & 5     & \phantom{0}45.6 & \phantom{0}57.2 & \phantom{0}68.5 & \phantom{0}81.4 & 100.0           &                 &                 &                 &       \\
		                             & 6     & \phantom{0}43.1 & \phantom{0}53.0 & \phantom{0}62.2 & \phantom{0}72.0 & \phantom{0}83.3 & 100.0           &                 &                 &       \\
		                             & 7     & \phantom{0}41.1 & \phantom{0}49.9 & \phantom{0}57.9 & \phantom{0}65.9 & \phantom{0}74.5 & \phantom{0}84.7 & 100.0           &                 &       \\
		                             & 8     & \phantom{0}39.5 & \phantom{0}47.5 & \phantom{0}54.6 & \phantom{0}61.5 & \phantom{0}68.6 & \phantom{0}76.4 & \phantom{0}85.7 & 100.0           &       \\
		                             & 9     & \phantom{0}38.3 & \phantom{0}45.7 & \phantom{0}52.1 & \phantom{0}58.2 & \phantom{0}64.4 & \phantom{0}70.9 & \phantom{0}78.1 & \phantom{0}86.6 & 100.0 \\
		\hline
		\multirow{9}{*}{$\beta=0.2$} & 1     & 100.0           &                 &                 &                 &                 &                 &                 &                 &       \\
		                             & 2     & \phantom{0}68.1 & 100.0           &                 &                 &                 &                 &                 &                 &       \\
		                             & 3     & \phantom{0}57.3 & \phantom{0}76.3 & 100.0           &                 &                 &                 &                 &                 &       \\
		                             & 4     & \phantom{0}51.6 & \phantom{0}65.8 & \phantom{0}80.5 & 100.0           &                 &                 &                 &                 &       \\
		                             & 5     & \phantom{0}47.8 & \phantom{0}59.6 & \phantom{0}70.8 & \phantom{0}83.1 & 100.0           &                 &                 &                 &       \\
		                             & 6     & \phantom{0}45.2 & \phantom{0}55.4 & \phantom{0}64.7 & \phantom{0}74.2 & \phantom{0}84.8 & 100.0           &                 &                 &       \\
		                             & 7     & \phantom{0}43.1 & \phantom{0}52.3 & \phantom{0}60.3 & \phantom{0}68.3 & \phantom{0}76.6 & \phantom{0}86.0 & 100.0           &                 &       \\
		                             & 8     & \phantom{0}41.5 & \phantom{0}49.8 & \phantom{0}57.0 & \phantom{0}63.9 & \phantom{0}70.9 & \phantom{0}78.4 & \phantom{0}86.9 & 100.0           &       \\
		                             & 9     & \phantom{0}40.2 & \phantom{0}47.9 & \phantom{0}54.4 & \phantom{0}60.6 & \phantom{0}66.7 & \phantom{0}73.0 & \phantom{0}79.8 & \phantom{0}87.7 & 100.0 \\
		\bottomrule                                   
	\end{tabular}%
	\caption{Information rates (\%) for the optimal O'Brien-Fleming design incorporating up to eight interim analyses for efficacy with a one-sided type I error rate of 0.025.}
	\label{tab:313}
\end{table}

\begin{figure}[!ht]
	\centering
	\includegraphics[width=\linewidth]{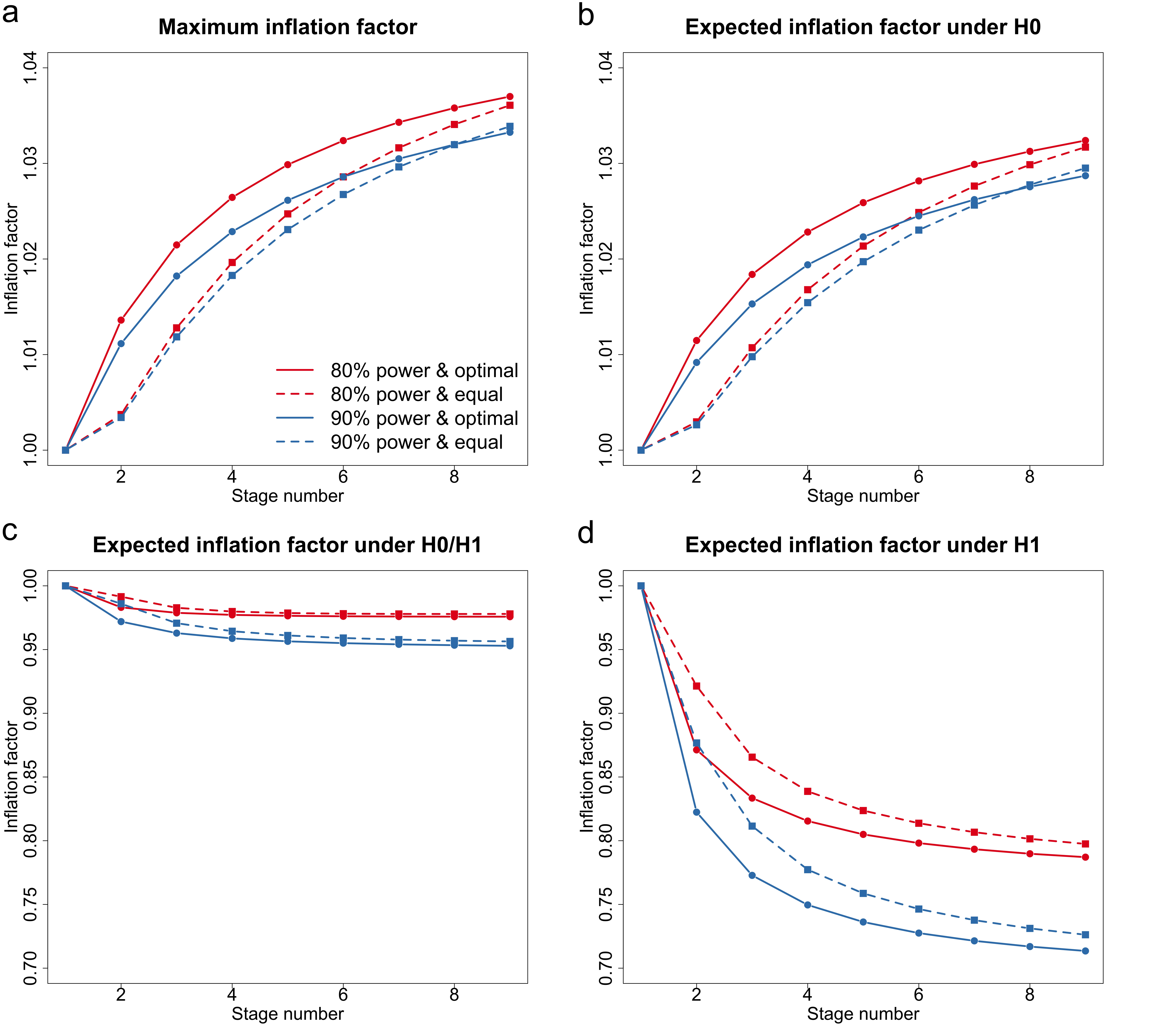}
	\caption{Operating characteristics for the optimal O'Brien-Fleming design incorporating up to eight interim analyses for efficacy with a one-sided type I error rate of 0.025.}
	\label{fig:313}
\end{figure}

Figure~\ref{fig:313} illustrates that, similar to the Pocock design, both the MIF and the EIF under $H_{0}$ rise sharply when moving from two to four total stages, after which the rate of increase slows, indicating diminishing inflation costs with additional stages. The overall magnitude of inflation, however, is markedly smaller in the O’Brien-Fleming design compared with the Pocock design. At lower stage counts, optimal spacing produces slightly higher inflation than equal scheduling, but this difference narrows as the number of stages increases and may even reverse at higher stage counts. Under $H_{0}/H_{1}$, as in the Haybittle-Peto design, inflation shows only a slight decline with additional stages, and the advantage of optimal spacing remains small and diminishes as interim analyses increase. Under $H_{1}$, inflation declines steeply as the number of stages increases, with the greatest drop occurring between two and four stages, after which the rate of decrease slows. The magnitude of this reduction falls between that observed in the Haybittle-Peto and Pocock designs. In this case, optimal scheduling consistently achieves lower inflation than equal spacing, although the incremental benefit declines with additional stages.

Unlike the Haybittle-Peto and Pocock designs, optimal scheduling in the O'Brien-Fleming design reduces inflation when a treatment effect exists, but at the cost of higher inflation when no treatment effect is present and a larger required maximum sample size. This advantage over conventional designs becomes more pronounced when a higher power is required. Furthermore, the greatest differences in both gains and costs between optimal and conventional designs occur with a single interim analysis, and these differences diminish as the number of interim analyses increases.

\subsection{The HYPRESS trial}
\label{sec:321}

We redesign the HYPRESS trial using the O’Brien-Fleming method, scheduling two interim analyses optimally at information rates of 57.4\% and 76.3\%, as determined with \texttt{OptimInterim} to minimise the ESS under the alternative hypothesis. Table~\ref{tab:321} summarises the probabilities of termination for efficacy at each interim analysis under the null hypothesis ($H_{0}:\theta_{p}=0.40,\theta_{h}=0.40$) and the alternative hypothesis ($H_{1}:\theta_{p}=0.40,\theta_{h}=0.25$), where $\theta_{p}$ and $\theta_{h}$ denote the 14-day occurrence of septic shock in the placebo group and the hydrocortisone group, respectively. Table~\ref{tab:322} presents the MSS and ESS’s under $H_{0}$, $H_{0}/H_{1}$ and $H_{1}$, with $H_{0}/H_{1}$ corresponding to the hypothesis that $\theta_{p}=0.40$ and $\theta_{h}=0.325$. Full design outputs for the HYPRESS trial using both the original and optimal interim analysis timings, generated through \texttt{rpact}, are available in Supplemental Material, Files~\href{run:./ZH2023_Supplemental_Material.pdf}{S1} and~\href{run:./ZH2023_Supplemental_Material.pdf}{S2}, respectively.

\begin{table}[!ht]
	\centering
	\begin{tabular}{ccccccc}
		\toprule
		      & \multicolumn{3}{c}{Original timing}                                 & \multicolumn{3}{c}{Optimal timing}                                  \\
		\cline{2-4}
		\cline{5-7}
		Stage & Sample size   & \multicolumn{2}{c}{$\Pr(\text{stop for efficacy})$} & Sample size   & \multicolumn{2}{c}{$\Pr(\text{stop for efficacy})$} \\
		\cline{3-4} 
		\cline{6-7}
		      &               & $H_{0}$ & $H_{1}$                                   &               & $H_{0}$ & $H_{1}$                                   \\
		\hline
		1     & 102.5 (0.333) & 0.0002  & 0.0186                                    & 177.9 (0.574) & 0.0062  & 0.2760                                    \\
		2     & 205.1 (0.667) & 0.0119  & 0.3988                                    & 236.8 (0.763) & 0.0144  & 0.2802                                    \\
		3     & 307.6 (1.000) & 0.0379  & 0.3826                                    & 310.3 (1.000) & 0.0294  & 0.2438                                    \\
		\bottomrule
	\end{tabular}%
	\caption{Probabilities of stopping for efficacy under $H_{0}$ and $H_{1}$ for the HYPRESS trial with the original timing and optimal timing of two interim analyses for efficacy. Figures in brackets are the information rates.}
	\label{tab:321}
\end{table}

\begin{table}[!ht]
	\centering
	\begin{tabular}{ccc}
		\toprule
		Sample size             & Original timing & Optimal timing \\
		\hline                                
		MSS                     & 307.6           & 310.3 	       \\
		ESS under $H_{0}$       & 306.4           & 308.4          \\
		ESS under $H_{0}/H_{1}$ & 298.5           & 297.3          \\
		ESS under $H_{1}$       & 262.9           & 253.1          \\
		\bottomrule                                   
	\end{tabular}%
	\caption{MSS and ESSs under $H_{0}$, $H_{0}/H_{1}$ and $H_{1}$ for the HYPRESS trial with the original timing and optimal timing of two interim analyses for efficacy.}
	\label{tab:322}
\end{table}

From Table~\ref{tab:321}, the optimal design, compared with the original design, requires a larger sample size at each interim and final analysis but markedly increases the probability of early stopping for efficacy (from 41.7\% to 55.6\%), with a particularly notable gain occurring at the first interim analysis (from 1.9\% to 27.6\%). When hydrocortisone decreases the 14-day occurrence of septic shock by 15\%, this higher probability of early stop translates into a 3.7\% reduction in the ESS under $H_{1}$, albeit with a modest 0.9\% increase in the MSS (see Table~\ref{tab:322}). Even in cases where the true treatment effect is absent or more modest than anticipated, the optimal design maintains comparable performance, with only a 0.7\% increase in the ESS under $H_{0}$ and a 0.4\% decrease under $H_{0}/H_{1}$. Overall, the optimal design for HYPRESS yields a clear average trial size saving when a clinically meaningful benefit from hydrocortisone exists, while incurring only a minimal increase in the maximum trial size.

\begin{figure}[!ht]
	\centering
	\includegraphics[width=\linewidth]{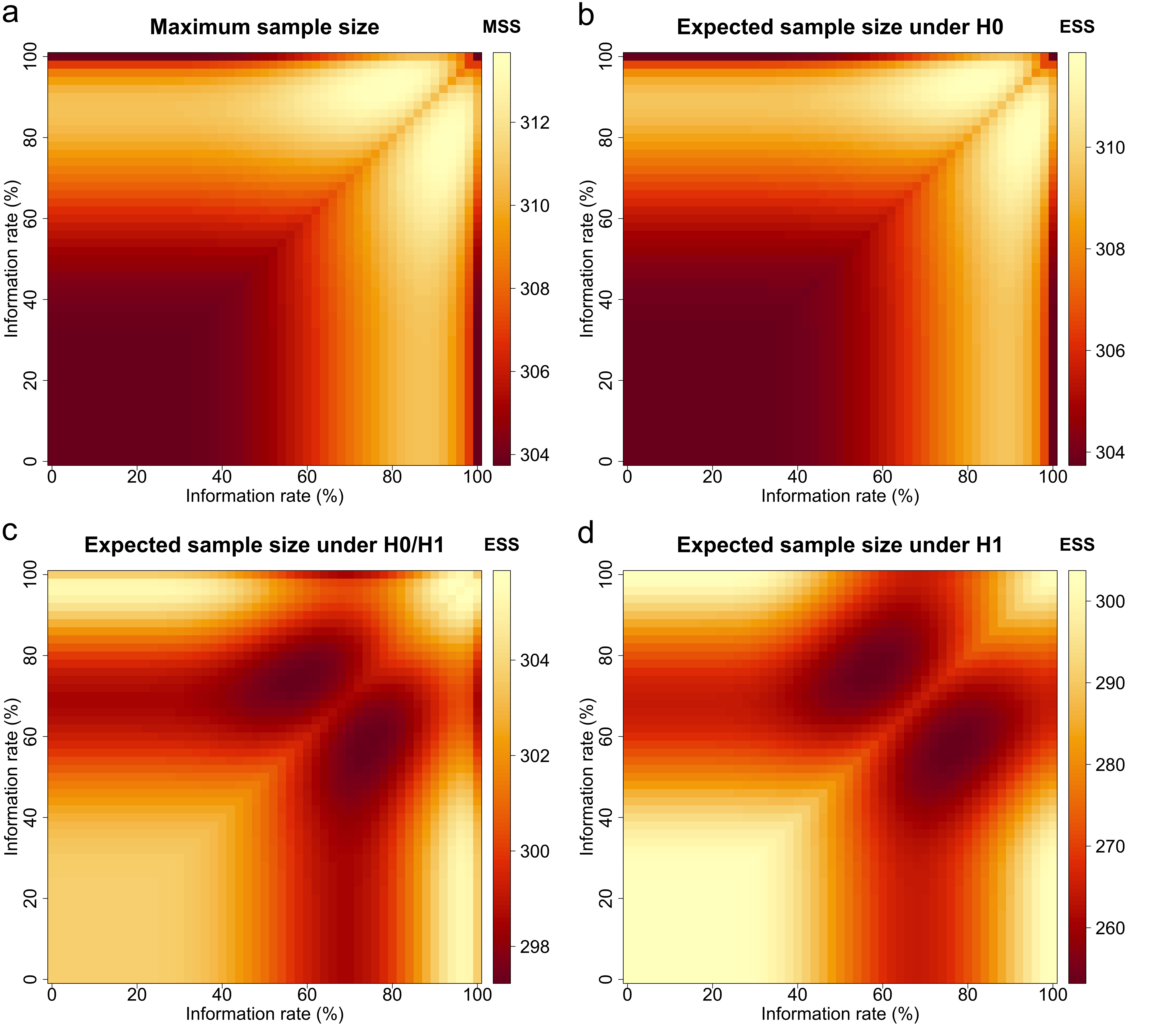}
	\caption{MSS and ESSs under $H_{0}$, $H_{0}/H_{1}$ and $H_{1}$ for the HYPRESS trial across all possible timing for up to two interim analyses for efficacy.}
	\label{fig:321}
\end{figure}

To further explore the operating characteristics of the optimal design for HYPRESS, we plot the MSS and ESSs under $H_{0}$, $H_{0}/H_{1}$ and $H_{1}$ across all possible timing for up to two interim analyses in Figure~\ref{fig:321}. To gain a smaller ESS under $H_{1}$, Figure~\ref{fig:321}d suggests planning both interim analyses between about 50\% and 90\% of the MSS with available primary outcome data, which however necessitates a larger ESS under $H_{0}$ (see Figure~\ref{fig:321}b). This conflict results in the optimal design penalising HYPRESS for accepting a larger MSS (see Figure~\ref{fig:321}a). Furthermore, as shown in Figures~\ref{fig:321}c and \ref{fig:321}d, the optimal timing of interim analyses nearly minimises the ESS under $H_{0}/H_{1}$, implying its effectiveness even in cases where the difference in true treatment effects is modest.

\subsection{The ADRENAL trial}
\label{sec:322}

We redesign the ADRENAL trial using the Haybittle-Peto approach, scheduling two interim analyses at optimally determined information rates of 44.4\% and 70.5\%. These timings, obtained by \texttt{OptimInterim}, are chosen to minimise the ESS under the alternative hypothesis. We present exit probabilities for efficacy at each interim analysis under both the null hypothesis ($H_{0}:\theta_{p}=0.33,\theta_{h}=0.33$) and the alternative hypothesis ($H_{1}:\theta_{p}=0.33,\theta_{h}=0.28$) in Table~\ref{tab:323}, where $\theta_{p}$ and $\theta_{h}$ represent the 90-day all-cause mortality in the placebo group and the hydrocortisone group, respectively. We outline the MSS and ESSs under $H_{0}$, $H_{0}/H_{1}$ and $H_{1}$ in Table~\ref{tab:324}, where $H_{0}/H_{1}$ is the hypothesis that $\theta_{p}=0.33$ and $\theta_{h}=0.305$. Comprehensive analytical outputs for the ADRENAL trial design using both the original interim analysis schedule and the optimised schedule generated via \texttt{rpact} are left in Supplemental Material, File~\href{run:./ZH2023_Supplemental_Material.pdf}{S3} and \href{run:./ZH2023_Supplemental_Material.pdf}{S4}, respectively.

\begin{table}[!ht]
	\centering
	\begin{tabular}{ccccccc}
		\toprule
		      & \multicolumn{3}{c}{Original timing}                                            & \multicolumn{3}{c}{Optimal timing}                                   \\
		\cline{2-4}
		\cline{5-7}
		Stage & Sample size              & \multicolumn{2}{c}{$\Pr(\text{stop for efficacy})$} & Sample size    & \multicolumn{2}{c}{$\Pr(\text{stop for efficacy})$} \\
		\cline{3-4} 
		\cline{6-7}
		      &                          & $H_{0}$ & $H_{1}$                                   &                & $H_{0}$ & $H_{1}$                                   \\
		\hline
		1     & \phantom{0}897.3 (0.250) & 0.0027  & 0.0849                                    & 1587.8 (0.444) & 0.0027  & 0.2018                                    \\
		2     & 2361.4 (0.658)           & 0.0024  & 0.2895                                    & 2518.2 (0.704) & 0.0020  & 0.2140                                    \\
		3     & 3589.4 (1.000)           & 0.0449  & 0.5256                                    & 3574.7 (1.000) & 0.0453  & 0.4842                                    \\
		\bottomrule
	\end{tabular}%
	\caption{Probabilities of stopping for efficacy under $H_{0}$ and $H_{1}$ for the ADRENAL trial with the original timing and optimal timing of two interim analyses for efficacy. Figures in brackets are the information rates.}
	\label{tab:323}
\end{table}

\begin{table}[!ht]
	\centering
	\begin{tabular}{ccc}
		\toprule
		Sample size             & Original timing & Optimal timing \\
		\hline                                
		MSS                     & 3589.4          & 3574.7	       \\
		ESS under $H_{0}$       & 3579.2          & 3567.2         \\
		ESS under $H_{0}/H_{1}$ & 3501.2          & 3483.8         \\
		ESS under $H_{1}$       & 3005.2          & 2947.6         \\
		\bottomrule                                   
	\end{tabular}%
	\caption{MSS and ESSs under $H_{0}$, $H_{0}/H_{1}$ and $H_{1}$ for the ADRENAL trial with the original timing and optimal timing of two interim analyses for efficacy.}
	\label{tab:324}
\end{table}

As demonstrated in Table~\ref{tab:323}, the optimal design schedules larger sample sizes at each interim analysis but a smaller sample size at the final analysis compared with the original design. This adjustment of information increases the overall probability of early stopping for efficacy by 4.1\% when hydrocortisone achieves a 5\% absolute reduction in all-cause mortality. The most notable gain occurs at the first interim analysis, where the probability of early stopping rises markedly from 8.5\% to 20.2\%. In terms of efficiency, the optimal design reduces the ESS by 1.9\% under $H_{1}$ and 0.5\% under $H_{0}/H_{1}$. Unlike the findings for HYPRESS, it also yields a 0.3\% decrease in the ESS under $H_{0}$ and a 0.5\% reduction in the MSS, as summarised in Table~\ref{tab:324}. Overall, across all considered treatment effect scenarios, the optimal design yields a lower average sample size requirement than the original, while simultaneously reducing the maximum trial size.

\begin{figure}[!ht]
	\centering
	\includegraphics[width=\linewidth]{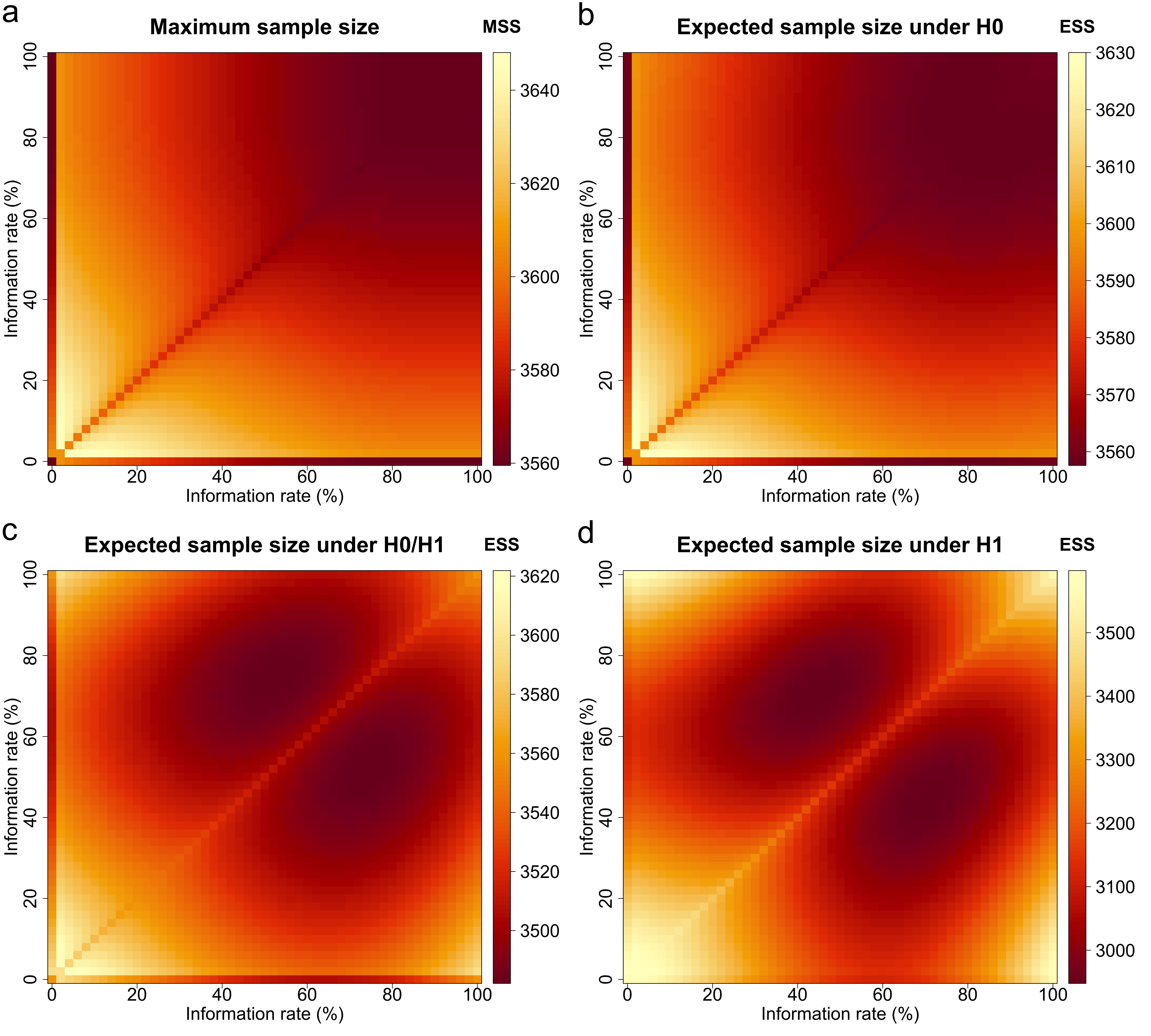}
	\caption{MSS and ESSs under $H_{0}$, $H_{0}/H_{1}$ and $H_{1}$ for the ADRENAL trial across all possible timing for up to two interim analyses for efficacy.}
	\label{fig:322}
\end{figure}

We present plots in Figure~\ref{fig:322} that demonstrates the MSS and ESSs under $H_{0}$, $H_{0}/H_{1}$ and $H_{1}$ across all possible timing for up to two interim analyses to further investigate the operating characteristic of the optimal design for ADRENAL. Like HYPRESS, we observe that the optimal timing of interim analyses nearly minimises the ESS under $H_{0}/H_{1}$ in ADRENAL (see Figures~\ref{fig:322}c and \ref{fig:322}d). Figures~\ref{fig:322}c and \ref{fig:322}d also suggest avoiding planning any interim analyses too early (before the information rate reaches approximately 30\%) or too late (after the information rate reaches about 90\%) but running at least one interim analysis after the information rate reaches around 60\% if hydrocortisone reduces the all-cause mortality. Combining with Figure~\ref{fig:322}b, which suggests planning both interim analyses after around 60\% of the MSS with available primary outcome data if there is no benefit of hydrocortisone, we find a certain overlap between the two strategies. This still leads to the optimal design for ADRENAL increasing the MSS similarly to HYPRESS but only to a slight extent (see Figure~\ref{fig:322}a).

As shown by \citet{li2023}, adding a futility boundary to the ADRENAL trial could have resulted in early termination for futility after enrolling 2926 of the planned 3658 patients, if the trial had been redesigned using the O’Brien-Fleming approach with four equally spaced interim analyses for both efficacy and futility. This reanalysis demonstrates that futility monitoring can further enhance the potential sample size savings achievable with GSDs. Using the Lan-DeMets error spending approach \citep{lan1983}, we investigated the operating characteristics of the optimal Pocock and O'Brien-Fleming designs with up to eight interim analyses incorporating both efficacy and (non-binding) futility, given a one-sided type I error rate of 0.025 and a type II error rate of 0.1 and 0.2. Full results, including the optimal interim analysis timings, stopping boundaries and operating characteristics, are provided in Supplemental Material, Tables~\href{run:./ZH2023_Supplemental_Material.pdf}{S4}--\href{run:./ZH2023_Supplemental_Material.pdf}{S7} and Figures~\href{run:./ZH2023_Supplemental_Material.pdf}{S4}--\href{run:./ZH2023_Supplemental_Material.pdf}{S7}.

\section{Discussion}
\label{sec:4}
In this work, we have introduced \texttt{OptimInterim} to address the challenge of determining the optimal timing of interim analyses in group sequential trials, aiming to minimise the ESS under a specified alternative hypothesis while maintaining control of type I and type II error rates. Our results have shown that GSDs incorporating optimally spaced interim analyses achieve desirable ESS reductions without compromising the MSS across a variety of endpoint types, effect sizes, error rates and stopping rules.
The practical application of \texttt{OptimInterim}, in combination with \texttt{rpact}, has been illustrated with the HYPRESS and ADRENAL trials, where optimally spaced interim analyses delivered clear efficiency gains compared with the original trial designs.
We have assessed the operating characteristics of three widely adopted GSDs, the Haybittle-Peto, Pocock and O'Brien-Fleming designs, incorporating up to eight interim analyses for efficacy. When optimally spaced interim analyses are applied, these designs can achieve substantial efficiency improvements, with reductions in the ESS under $H_{1}$ ranging from at least 8.0\% to as much as 34.4\% compared to a fixed design. These gains, however, are accompanied by increases in the MSS, reflecting the trade-off between early stopping potential and maximum trial size. While ESS reductions tend to increase as more interim analyses are added, the marginal benefits diminish substantially after three interim looks, with any additional analysis beyond this point producing at most a 1.8\% further decrease in the ESS under $H_{1}$. For example, in the HYPRESS trial employing the O'Brien-Fleming design, scheduling three optimally spaced interim analyses results in an ESS of 247.7 under $H_{1}$, while adding a fourth interim analysis achieves a negligible additional saving of just 3.2 patients. These findings provide practical guidance for determining both the timing and the number of interim analyses, supporting designs that balance efficiency gains with operational complexity.

Overall, the optimal Haybittle-Peto and Pocock designs yield only modest additional sample size savings, with additional reductions of up to 1.2\% in both MSS and ESSs across $H_{0}$, $H_{0}/H_{1}$ and $H_{1}$ when compared with those incorporating equally spaced interim analyses. This is largely because alpha spending is distributed relatively evenly across interim analyses in these designs, meaning that equally spaced timings are already close to optimal. The optimal O’Brien-Fleming design, in contrast, delivers a more substantial improvement in sample size efficiency under $H_{1}$, with the ESS reduced by as much as 5.4\% compared with the conventional version. This gain is accompanied by a modest increase in the MSS, up to 1.0\%, which diminishes, or even disappears, as the number of interim analyses grows. Importantly, the optimal O’Brien-Fleming design can match the performance of the conventional design while using one fewer interim analysis, offering a favourable balance between efficiency and operational simplicity. In summary, while optimally spaced interim analyses yield only marginal benefits in the Haybittle-Peto and Pocock designs, they achieve meaningful ESS reductions in the O’Brien-Fleming design, with the trade-off being a minimal and often manageable increase in MSS.

Compared to the optimal Pocock and O’Brien-Fleming designs that include only an efficacy stopping boundary, adding a futility boundary yields additional ESS reductions of at least 25.9\% under $H_{0}$ and 8.8\% under $H_{0}/H_{1}$. These reductions arise because the futility boundary allows trials with little or no treatment effect to stop earlier, thereby avoiding unnecessary recruitment and follow-up. However, these benefits come at the cost of a substantial increase in the MSS (\textit{i.e.}, up to 29.0\% for the optimal Pocock design and 14.6\% for the optimal O’Brien-Fleming design). This trade-off reflects the more conservative structure of GSDs with dual stopping boundaries, which necessitate a larger MSS to maintain nominal type I and type II error rates while allowing for early termination in both directions. The decision to include a futility boundary should thus be informed by the relative importance of minimising the average trial size versus constraining the maximum trial size in a given clinical and operational context.

Importantly, our findings have shown that for given type I and type II error rates and stopping rules, the operating characteristics are preserved, and the optimal interim analysis timings remain unchanged across different effect sizes, regardless of whether the endpoint is continuous or binary. Based on this invariance, we provide reference tables summarising optimal timings for up to eight interim analyses in group sequential trials with commonly adopted type I and type II error rates and stopping boundaries (see Tables~\ref{tab:311}--\ref{tab:313} and Supplemental Material, Tables~\href{run:./ZH2023_Supplemental_Material.pdf}{S4} and \href{run:./ZH2023_Supplemental_Material.pdf}{S6}), as well as their operating characteristics (see Supplemental Material, Tables~\href{run:./ZH2023_Supplemental_Material.pdf}{S1}--\href{run:./ZH2023_Supplemental_Material.pdf}{S3}, \href{run:./ZH2023_Supplemental_Material.pdf}{S5} and \href{run:./ZH2023_Supplemental_Material.pdf}{S7}). 
These tables therefore offer a ready-to-use guide for determining both the number and timing of interim analyses in GSDs. Notably, the optimal timing of interim analyses is identical for a one-sided type I error rate of 0.025 and a two-sided type I error rate of 0.05. 

\texttt{OptimInterim} has been demonstrated to offer the prospect of increased statistical efficiency in group sequential trials with continuous or binary endpoints, providing the flexibility of incorporating multiple interim analyses and various stopping rules, but there remain several potential extensions. For example, \texttt{OptimInterim} cannot currently be applied directly to group sequential trials with survival endpoints. Because the log-rank test, \textit{i.e.}, the standard test statistic for survival data, is asymptotically normally distributed, extending \texttt{OptimInterim} to this setting is conceptually straightforward. However, survival endpoints introduce additional complexities resulting from follow-up time and study duration, which necessitate a thorough reexamination of the operating characteristics of the incorporating optimally spaced interim analyses for survival endpoints.

\texttt{OptimInterim} minimises the ESS under $H_{1}$, as this optimality criterion reflects the efficiency gain of a GSD when the trial is appropriately stopped early for efficacy, rejecting $H_{0}$.
However, in practice, the true effect size may be smaller than that assumed under $H_{1}$. In such scenarios, a weighted combination of ESS values for selected effect sizes is usually used as the optimality criterion \citep[\textit{e.g.},][where $\sum_{i=0}^{1}\mathbb{E}(N \mid H_{i})/2$ was explored]{xi2017}. A further generalisation is to integrate the ESS over a distribution of plausible effect sizes \citep[\textit{e.g.},][]{lokhnygina2008},
\begin{linenomath}
	\begin{equation*}
		\mathbb{E}(N)
		=
		\int_{\Delta} \mathbb{E}(N \mid \delta) \,d\rho(\delta),
	\end{equation*}
\end{linenomath}
where $\rho$ denotes a probability measure on the sample space $\Delta$ that captures the prior uncertainty about the effect size. Note that $\mathbb{E}(N)=\mathbb{E}(N \mid H_{0})$ if we take $\rho$ to be a Dirac measure centred on $\delta_{0}$, and $\mathbb{E}(N)=\mathbb{E}(N \mid H_{1})$ if we take $\rho$ to be a Dirac measure centred on $\delta_{1}$. As demonstrated in this work, minimising the ESS can result in a larger MSS. An alternative extension can involve minimising the weighted combination of two or more optimality criteria of interest, such as the ESS, MSS and standard deviation of the sample size \citep{grayling2021}, resulting in a more balanced design.

\texttt{OptimInterim} optimises the scheduling of interim analyses with predefined stopping rules, as in practice, group sequential trials are commonly planned with the Haybittle-Peto, Pocock and O'Brien-Fleming decision rules (\textit{e.g.}, used in over 60\% of cardiovascular group sequential trials \citep{zhang2023}). Like \citet{brittain1993}, \texttt{OptimInterim} can be extended to jointly optimise stopping boundaries and interim look timings, which is expected to further improve statistical efficiency but at the cost of increasing computational complexity, in particular with an increase in the number of interim analyses. To circumvent this challenge, as in \citet{anderson2007} and \cite{wason2015}, we can characterise stopping boundaries through the power family \citep{emerson1989,pampallona1994,wason2015}, \textit{i.e.}, the shape of the stopping boundaries is fully controlled by one or two parameters. The power family stopping boundaries provide great flexibility in possible shapes while largely reducing the search space, which results in a substantial reduced computational cost in optimisation. Other family stopping boundaries can be found in \textit{e.g.}, \citet{wang1987}, \citet{jennison1987} and \citet{hwang1990}.

Moreover, \texttt{OptimInterim} assumes that treatment outcomes are immediately observed while subject enrolment is halted at the interim analysis, which is not common in clinical trial practice. As in \citet{togo2013}, we can incorporate follow-up duration and allow subject enrolment to continue at the interim analysis. Consequently, the number of subjects enrolled is no longer the same as that of subjects observed. Let $r(\tau)$ denote the enrolment rate at the calendar time $\tau$, and the ESS given $\delta$ can then be reformulated as
\begin{linenomath}
	\begin{equation}
		\label{eqn:4102}
		\mathbb{E}(N \mid \delta)
		=
		N_{K} \biggl(t'_{1} + \sum_{k=2}^{K}(t'_{k}-t'_{k-1})F(\mathcal{C}_{1,1} \times \mathcal{C}_{1,2} \times \cdots \times \mathcal{C}_{1,k-1} \mid \boldsymbol{\mu}_{1:k},\boldsymbol{\Sigma}_{k \times k}) \biggr),
	\end{equation}
\end{linenomath}
where
\begin{linenomath}
	\begin{equation*}
		t'_{k}
		=
		\min\biggl(t_{k}+\frac{\int_{\tau_{k}}^{\tau_{k}+d}r(\tau) \,d\tau}{N_{K}}, 1\biggr)
	\end{equation*}
\end{linenomath}
with $\tau_{k}$ being the calendar time for the $k$-th interim analysis and $d$ being the follow-up duration. From Eq.~(\ref{eqn:4102}), we see that the optimal interim analysis timing in terms of minimising $\mathbb{E}(N \mid \delta)$ in Eq.~(\ref{eqn:4102}) largely relies on the follow-up duration and enrolment model. The reduction in the ESS achieved by incorporating the optimal timing of interim analyses diminishes with an increase in the follow-up duration and/or enrolment rate. The optimal timing of interim analyses in terms of minimising $\mathbb{E}(N \mid \delta)$ in Eq.~(\ref{eqn:4102}) remains optimal if the enrolment rate is fixed over time, and the follow-up duration and/or the enrolment rate are sufficiently small such that $t'_{k}<1$ holds for $k<K$. Refer to \citet{togo2013} for an investigation into two-stage group sequential trials.

In summary, \texttt{OptimInterim} provides a valuable resource for planning interim analyses across a wide range of group sequential trials. By offering a practical approach for determining optimal interim analysis timings, it has the potential to enhance the efficiency gains achievable with such designs. Importantly, since the optimal scheduling depends only on the specified type I and type II error rates and stopping rules and not on the endpoint types or effect sizes, the accompanying reference tables are broadly generalisable. As such, they provide readily applicable guidance for a broad range of GSDs, supporting more efficient and well-informed decision-making in clinical research.
\subsection*{Financial disclosure}
The authors declare no funding associated with the work presented in this article. 

\subsection*{Conflict of interest}
The authors declare no potential conflict of interests.






\bibliography{ZH2023_Manuscript}

\end{document}